\documentclass[mathleft
]{an}
\usepackage{graphicx}
\usepackage{times}
\begin{document}

\Pagespan{789}{}
\Yearpublication{2006}%
\Yearsubmission{2005}%
\Month{11}%
\Volume{999}%
\Issue{88}%

\title{Cosmic X-ray surveys of Active Galactic Nuclei: the synergy
  between X-ray and infrared observations}

\author{D.~M.~Alexander\inst{1}\fnmsep\thanks{Corresponding author:
  \email{d.m.alexander@durham.ac.uk}\newline}
}
\titlerunning{Synergy between X-ray and infrared observations}
\authorrunning{D.~M.~Alexander}
\institute{
$^1$ Centre for Extragalactic Astronomy,
Department of Physics,
Durham University,
South Road,
Durham,
DH1 3LE, UK}

\received{DD MMM 2016}
\accepted{DD MMM 2016}
\publonline{later}


\abstract{We briefly review the synergy between X-ray and infrared
  observations for Active Galactic Nuclei (AGNs) detected in cosmic
  X-ray surveys, primarily with {\it XMM-Newton}, {\it Chandra}, and
  {\it NuSTAR}. We focus on two complementary aspects of this
  X-ray--infrared synergy (1) the identification of the most heavily
  obscured AGNs and (2) the connection between star formation and AGN
  activity. We also briefly discuss future prospects for
  X-ray--infrared studies over the next decade.}

\maketitle

\section{Introduction}

There is great synergy between X-ray and thermal infrared observations
for the study of Active Galactic Nuclei (AGNs) and their connection to
star formation (SF).\footnote{In this paper the thermal infrared
  waveband is defined as 3--1000~$\mu$m.}

X-ray observations provide, arguably, the most efficient method
currently available for reliably identifying AGNs. This is due to a
combination of (1) the penetrating nature of X-rays, which allow even
obscured AGNs to be directly identified at X-ray energies and (2) the
low dilution from the host galaxy (i.e.,\ X-ray emission assocated to
stellar processes) at X-ray energies. X-ray surveys of AGNs therefore
provide one of the most complete and reliable approaches to construct
a cosmic census of AGN activity (see Brandt \& Alexander 2015 for a
recent review). However, the most heavily obscured AGNs will be weak
or undetected in X-rays, making their X-ray identification challenging
or impossible. Furthermore, since the X-ray emission from the host
galaxy is weak, X-ray surveys are not an effective method for
providing a cosmic census of SF. Fortunately, these short comings can
be mostly alleviated from a combination of thermal infrared and X-ray
observations.

The thermal infrared waveband provides an effective method to study
AGNs and their connection with SF (e.g.,\ Alexander \& Hickox 2012;
Lutz 2014). The dust in the obscuring medium, which makes the heavily
obscured AGNs weak or invisible at X-ray energies, will be heated by
the AGN and emitted thermally at infrared wavelengths. Heavily
obscured AGNs can therefore, in principle, be identified at infrared
wavelengths. However, since a significant fraction of SF is also
emitted at infrared wavelengths, due to the majority of stars forming
in dusty molecular clouds (see Kennuicutt \& Evans 2012 for a review),
the identification of AGNs in the infrared waveband is often not a
straightforward exercise. Fortunately, the AGN and SF components can
often be distinguished in the infrared waveband since the AGN
component has a ``hotter'' infrared energy distribution (SED;
i.e.,\ the AGN component is bright at mid-infrared wavelengths of
$\approx$~3--30~$\mu$m) and produces a harder ionisation radiation
field than the SF component (e.g.,\ Donley et~al. 2012; Mateos
et~al. 2012; Stern et~al. 2012; Assef et~al. 2013). Therefore, several
methods allow the infrared emission from the AGN component to be
reliably measured even in the presence of SF at infrared wavelengths:
(1) the decomposition of the SED into the AGN and SF components by
fitting AGN and SF galaxy templates to the infrared emission, (2)
high-spatial resolution infrared emission to isolate the AGN component
at the centre of the galaxy from the SF emission (currently only
possible for nearby AGNs), and (3) the detection of AGN and SF spectral
``signatures'' (i.e.,\ emission lines and solid-state features) using
infrared spectroscopy. Once the contribution from the AGN can be
constrained at infrared wavelengths, the thermal infrared is
arguably the most effective waveband for constructing the cosmic
census of SF (e.g.,\ Lutz 2014; Madau \& Dickinson 2014).

The combination of X-ray and thermal infrared observations therefore
provides the opportunity for a more complete understanding and cosmic
census of AGN activity and the exploration and measurement of the
connection between AGN activity and SF. In this short paper we briefly
review recent results on the X-ray--infrared properties of AGNs and
the identification of heavily obscured AGNs (see \S2) and the
connection between AGN activity and SF (see \S3). We focus on results
that use the most sensitive X-ray ({\it XMM-Newton} and {\it Chandra}
at $<$~10~keV and {\it NuSTAR} at $>$~10~keV; e.g.,\ Jansen
et~al. 2001; Weisskopf et~al. 2002; Harrison et~al. 2013) and infrared
({\it Herschel}; ALMA; ground-based mid-infrared observatories:
e.g.,\ Pilbratt et~al. 2010) observations available. In \S4 we briefly
review future propsects over the next decade.


\section{X-ray and infrared synergy: AGN properties and the 
identification of heavily obscured AGNs}

\begin{figure}
\centerline{
\includegraphics[width=75mm]{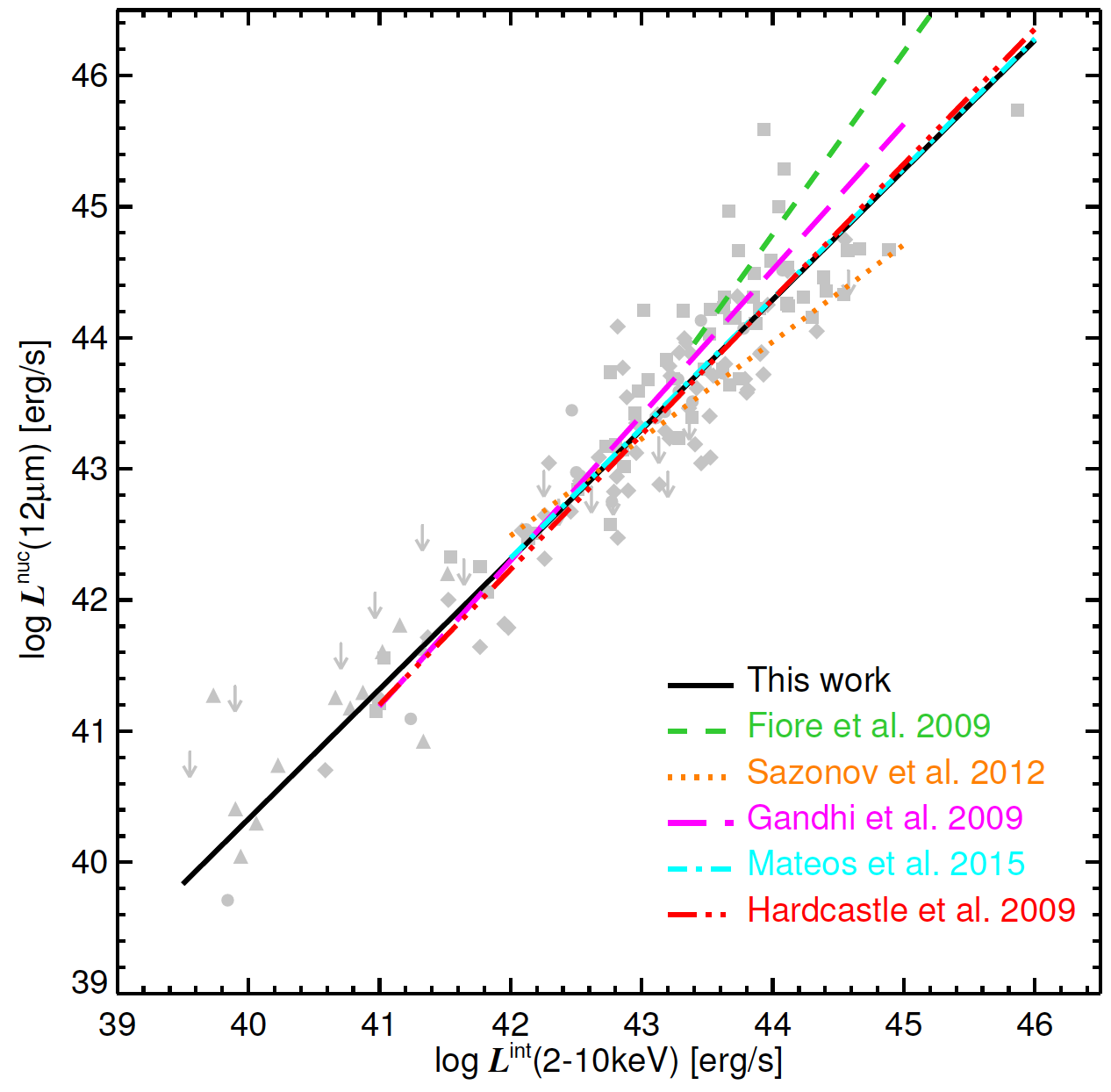}}
\caption{Observed mid-infrared luminosity versus rest-frame 2--10~keV
  luminosity, demonstrating the intrinsic mid-infrared--X-ray
  relationship for nearby AGNs. The 2--10~keV luminosity has been
  corrected for the presence of absorption through detailed X-ray
  spectral modelling of the X-ray data and the mid-infrared
  observations are from high-spatial resolution 12~$\mu$m imaging at
  the location of the AGN. The black line shows the best-fitting
  relationship from Asmus et~al. (2015) and the other lines show the
  best-fitting relationships from previous work (Fiore et~al. 2009;
  Gandhi et~al. 2009; Hardcastle et~al. 2009; Sazonov et~al. 2012;
  Mateos et~al. 2015). Taken from Asmus et~al. (2015).}
\label{label1}
\end{figure}

\begin{figure*}
\centerline{
\includegraphics[width=180mm]{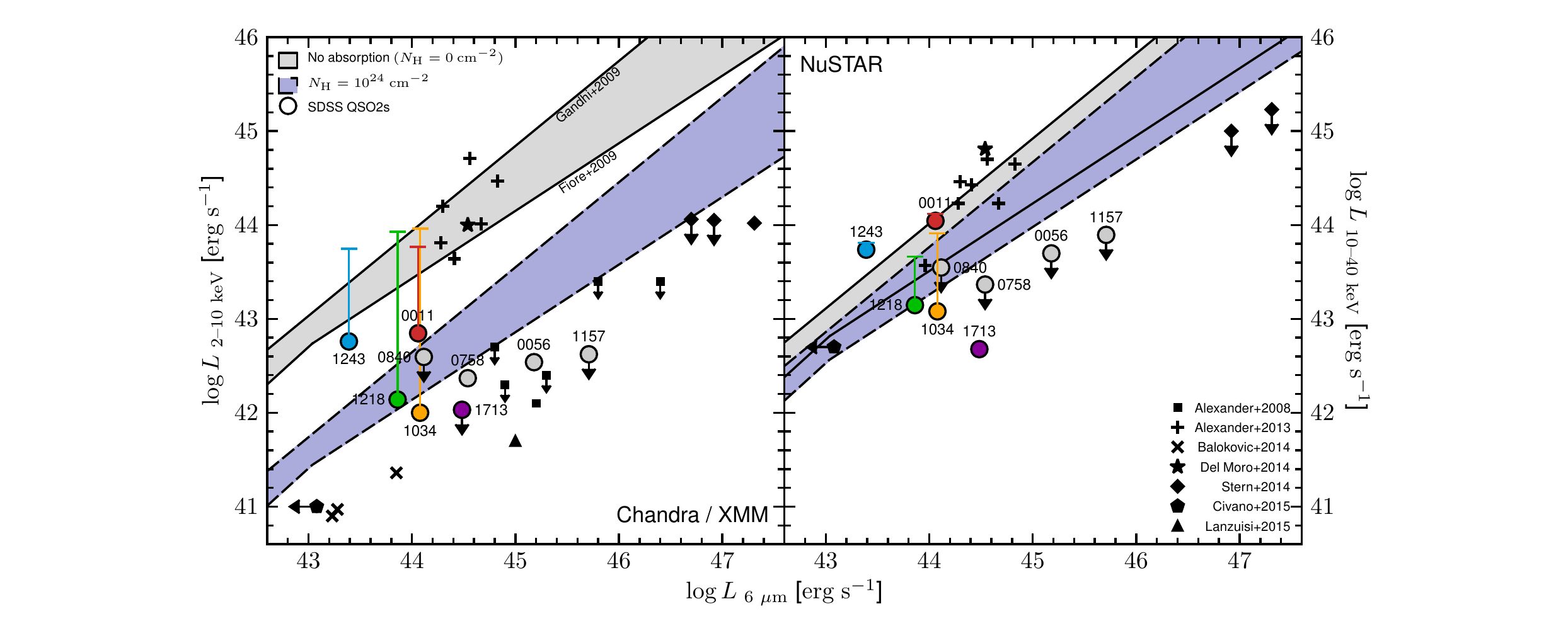}}
\caption{Observed X-ray luminosity versus rest-frame 6~$\mu$m
  luminosity ($\nu$$L_{\rm 6\mu m}$) in both the rest-frame 2--10~keV
  (left; from {\it Chandra} and {\it XMM-Newton}) and rest-frame
  10--40~keV (right; from {\it NuSTAR}) bands for a wide variety of
  AGNs (see Lansbury et~al. 2015 for more details). The solid curves
  indicate the expected intrinsic (i.e.,\ corrected for the presence
  of absorption) $L_{\rm X}$--$L_{\rm 6\mu m}$ relationship from two
  different studies (Gandhi et~al. 2009; Fiore et~al. 2009) while the
  grey shaded region indicates the difference in expectations between
  these two studies. The dashed curves and solid purple region show
  the effect of Compton-thick absorption ($N_{\rm
    H}=10^{24}$~cm$^{-2}$) on the observed X-ray emission. Taken from
  Lansbury et~al. (2015).}
\label{label1}
\end{figure*}

The combination of X-ray and thermal infrared observations provide
insight on the properties of the AGN, in addition to allowing for a
more complete census of AGNs over that obtained from just one of these
wavebands. The X-ray emission is predominantly produced in the
``corona'' and is thought to be due to inverse Compton scattering of
the primarily AGN accretion disc emission (which is produced mostly at
ultra-violet--optical wavelengths) by electrons above the accretion
disc (see Done 2010 for a review). By comparison, the infrared
emission is predominantly due to heated dust grains in the obscuring
structure (commonly referred to as the ``torus'' to indicate that it
is geometrically thick) and is essentially the cooler outer regions of
the accretion disc, where dust and molecules can form (see Elitzur \&
Shlosman 2006 for a review). Significant infrared emission is also
produced in the narrow-line region of some AGNs, perpendicular to the
accretion disc (e.g.,\ Tristram et~al. 2014; L{\'o}pez-Gonzaga
et~al. 2016). Given the different locations and sizes of the X-ray
corona and infrared ``torus'', the study of the X-ray and infrared
emission from AGNs gives insight on the relative geometry of the
corona--torus system (e.g.,\ the covering factor of the obscuring
``torus'').

In Fig.~1 we show the relationship between the observed 12~$\mu$m
luminosity and the intrinsic (i.e., corrected for the presence of
absorption) 2--10~keV luminosity for nearby AGNs. This plot is taken
from Asmus et~al. (2015), although the basic results are consistent
with previous works but with significantly improved source
statistics. The tightness of the infrared--X-ray luminosity
relationship indicates that the corona-torus geometry is broadly
constant and provides insight on the physical structure of the AGN. A
corrollary of this result is that the infrared emission from the AGN
can be used to predict the intrinsic X-ray emission even when the AGN
is so heavily obscured that it is weak or undetected at X-ray
energies.\footnote{The infrared emission is often not heavily obscured
  and so typically does not need to be corrected for obscuration.}
Consequently, the combination of X-ray and infrared observations can
be used to identify the most heavily obscured AGNs known, including
AGNs that are obscured by Compton-thick levels of absorption
(i.e.,\ when the absorbing column density is the inverse of the
Thomson scattering cross section: $N_{\rm H}>1/{\sigma_{\rm
    T}}>1.5\times10^{24}$~cm$^{-2}$).

Many studies have combined infrared and X-ray observations of AGNs to
identify the presence of Compton-thick absorbing column densities
(e.g.,\ Alexander et~al. 2008; Bauer et~al. 2010; Georgantopoulos
et~al. 2013). An example study is given in Fig.~2, which shows X-ray
luminosity versus rest-frame 6~$\mu$m luminosity for a variety of AGNs
studied from a combination of {\it XMM-Newton}, {\it Chandra}, and
{\it NuSTAR} observations. In the left-hand plot the X-ray luminosity
is at rest-frame 2--10~keV from {\it XMM-Newton} or {\it Chandra}
while in the right-hand plot the X-ray luminosity is at rest-frame
10--40~keV from {\it NuSTAR}. The two main results from these plots
are (1) there is a population of AGNs that are X-ray weak given their
mid-infrared luminosity and (2) the X-ray ``weakness'' of these AGNs
is less pronounced at 10--40~keV than at 2--10~keV. This X-ray
weakness is likely to be due to extreme levels of absorption, which
will effect the 2--10~keV emission more significantly than the
10--40~keV emission (e.g.,\ Wilms et~al. 2000), as demonstrated for
several sources from detailed X-ray spectral analysis which revealed
Compton-thick or near Compton-thick absorption (Lansbury
et~al. 2015). Several of the AGNs are even undetected in the X-ray
band and are presumably heavily Compton thick, although deeper X-ray
observations and further analyses would be required to confirm
this. We caution that, although mid-infrared--X-ray analyses provide
evidence for Compton-thick absorption, detailed X-ray spectral
analyses are required to provide definitive proof
(e.g.,\ Georgantopoulos et~al. 2011; Mateos et~al. 2015; Del Moro
et~al. 2016).

As may be expected, our census of AGN activity is less complete in the
distant Universe than found locally. However, surprisingly, we still
lack a complete census of even luminous AGN activity within our own
cosmic back yard ($d<15$~Mpc)! An example is the AGN in NGC~1448,
which lies at a distance of just $\approx$~12~Mpc but was only
discovered when observed with sensitive mid-infrared spectroscopy,
which revealed the presence of a high-excitation [NeV]~14.3~$\mu$m
emission line, the smoking gun evidence of AGN activity (Goulding \&
Alexander 2009). From the combination of {\it XMM-Newton}, {\it
  Chandra}, and {\it NuSTAR} X-ray spectral fitting with high-spatial
resolution ground-based mid-infrared imaging it is now clear that the
AGN in NGC~1448 is Compton thick, with a column density of $N_{\rm
  H}>5\times10^{24}$~cm$^{-2}$ (Annuar et~al. submitted). Indeed, it
seems likely that up-to half of the AGN population in the local
Universe is Compton thick (e.g.,\ Burlon et~al. 2011; Ricci
et~al. 2015). The fraction of the AGN population that are Compton
thick in the distant Universe is significantly more uncertain;
however, current results suggest that the fraction is broadly
comparable with that seen locally, at least for AGNs of comparable
luminosity (e.g.,\ Aird et~al. 2015; Buchner et~al. 2015; Lanzuisi
et~al. 2015; Del Moro et~al. 2016).

\section{X-ray and infrared synergy: the connection between AGN activity 
and star formation}

\begin{figure*}
\centerline{
\includegraphics[width=120mm]{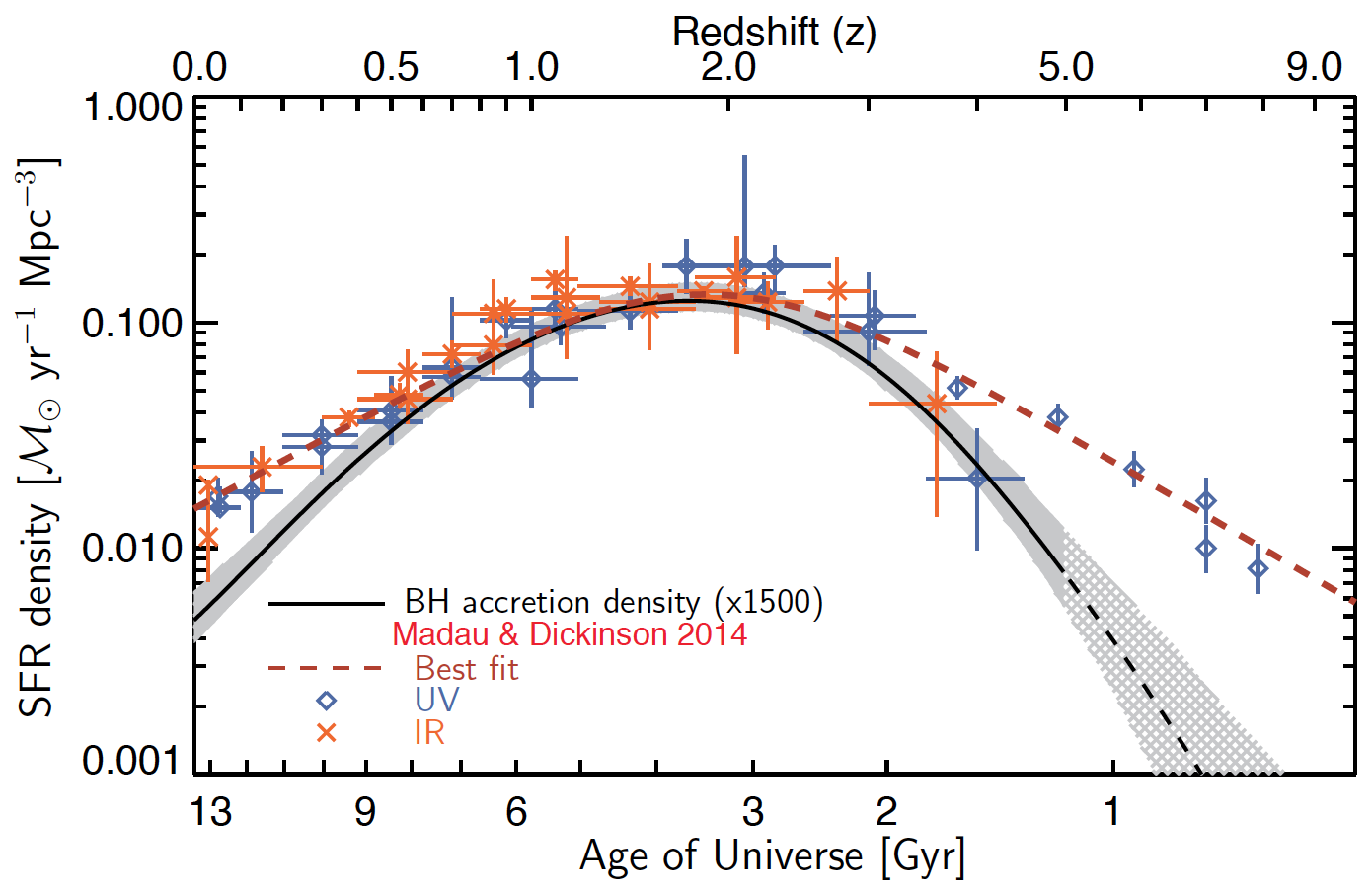}}
\caption{SF rate density per unit volume versus cosmic time for
  galaxies, using either infrared or ultra-violet measurements (as
  indicated); the best-fitting relationship is indicated by a dashed
  curve. The accretion rate density from X-ray AGNs is also shown as a
  black curve (uncertainities indicated by the grey shaded region) but
  scaled up by a factor of 1500. Taken from Aird et~al. (2015) using
  SF data from Madau \& Dickinson (2014).}
\label{label1}
\end{figure*}

The combination of X-ray and infrared observations allow for the study
of the connection between AGN activity and SF (e.g.,\ from the
measurement of AGN and SF activity in the X-ray and infrared bands,
respectively). To first order we should expect a connection between
AGN activity and SF since both processes are driven by essentially the
same gas supply: cold gas. SF is most efficient in cold dense
molecular clouds, where gravity can overcome the kinetic energy of the
gas and cause regions within the cloud to collapse and form stars. The
gas that resides in the AGN accretion disc, with a temperature of
$T\approx10^5$~K is not typically considered cold! However, the
majority of the gas would have originally been cold gas in the host
galaxy and driven into the vicinity of the accretion disc (e.g.,\ from
the loss of angular momentum and/or stellar/supernovae winds and
outflows; Jogee 2006) and then heated to high temperatures from
viscosity in the accretion disc.

Tight relationships between the properties of the galaxy and the
central supermassive black hole (e.g.,\ Kormendy \& Ho 2013; Graham
2016) potentially suggest a more symbiotic AGN--SF connection, whereby
the growth of the black hole and the galaxy are somehow regulated
(e.g.,\ due to inflows or outflows from the galaxy; Alexander \&
Hickox 2012; Fabian 2012; King \& Pounds 2015). Indeed, as shown in
Fig.~3, the cosmological evolution in SF rate density and accretion
rate density are remarkably similar, with an offset of a factor of
$\approx$~1500, which is broadly the average ratio between the mass of
the black hole and the galaxy spheroid for local galaxies. These
results provide evidence for a global AGN--SF connection; however,
they yield limited clues on the nature of the AGN--SF connection, such
as whether there are dependencies on the AGN and SF luminosity.

Prior to the launch of {\it Herschel} in 2010, studies on the SF
properties of distant X-ray AGNs were restricted to relatively low
sensitivity instruments or observations that were not optimal for
reliable measurements of the SF properties (e.g.,\ see \S5.5 of
Alexander \& Hickox 2012). {\it Herschel}, and more recently ALMA,
have greatly pushed forward our understanding of the AGN--SF
connection by providing sensitive observations at far-infrared
wavelengths, where the emission from dust-obscured SF peaks. From a
suite of studies performed with {\it Herschel} (often including
mid-infrared data from {\it Spitzer}) several key results on the SF
properties of X-ray AGNs have crystalised (e.g.,\ Harrison
et~al. 2012; Mullaney et~al. 2012a,b; Chen et~al. 2013; Rosario
et~al. 2013; DelVecchio et~al. 2014; Stanley et~al. 2015): (1) the
mean SF luminosity and the mean specific SF rate (i.e.,\ the stellar
mass divided by the SF rate, a measure of how quickly the galaxy is
growing) and their evolution with redshift are consistent with the SF
galaxy population, (2) the mean SF luminosity of X-ray AGNs is flat
with AGN luminosity, and (3) the mean AGN luminosity of
infrared-selected SF galaxies correlates with the SF luminosity.

\begin{figure*}
\centerline{
\includegraphics[width=100mm]{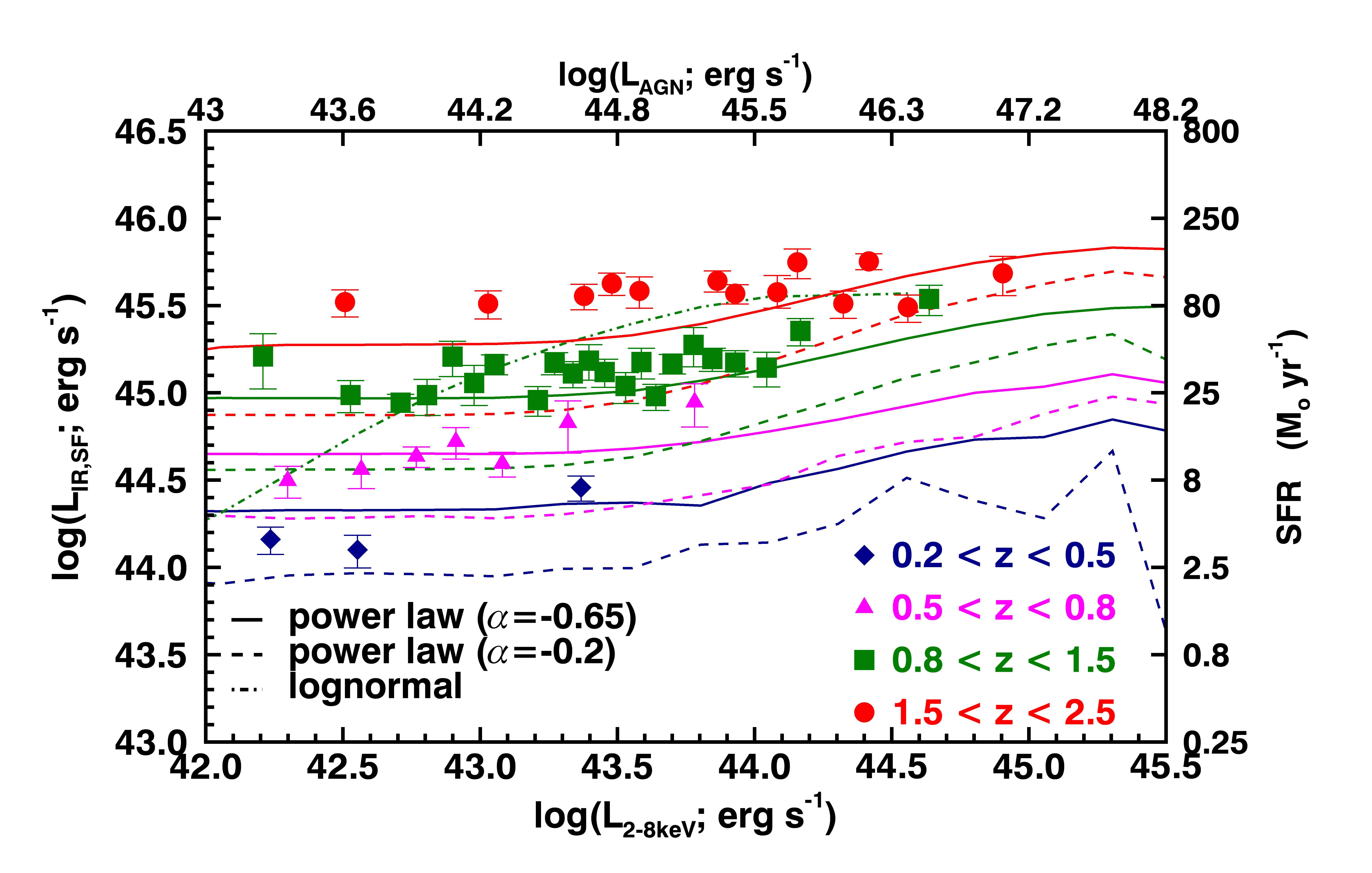}
\includegraphics[width=80mm]{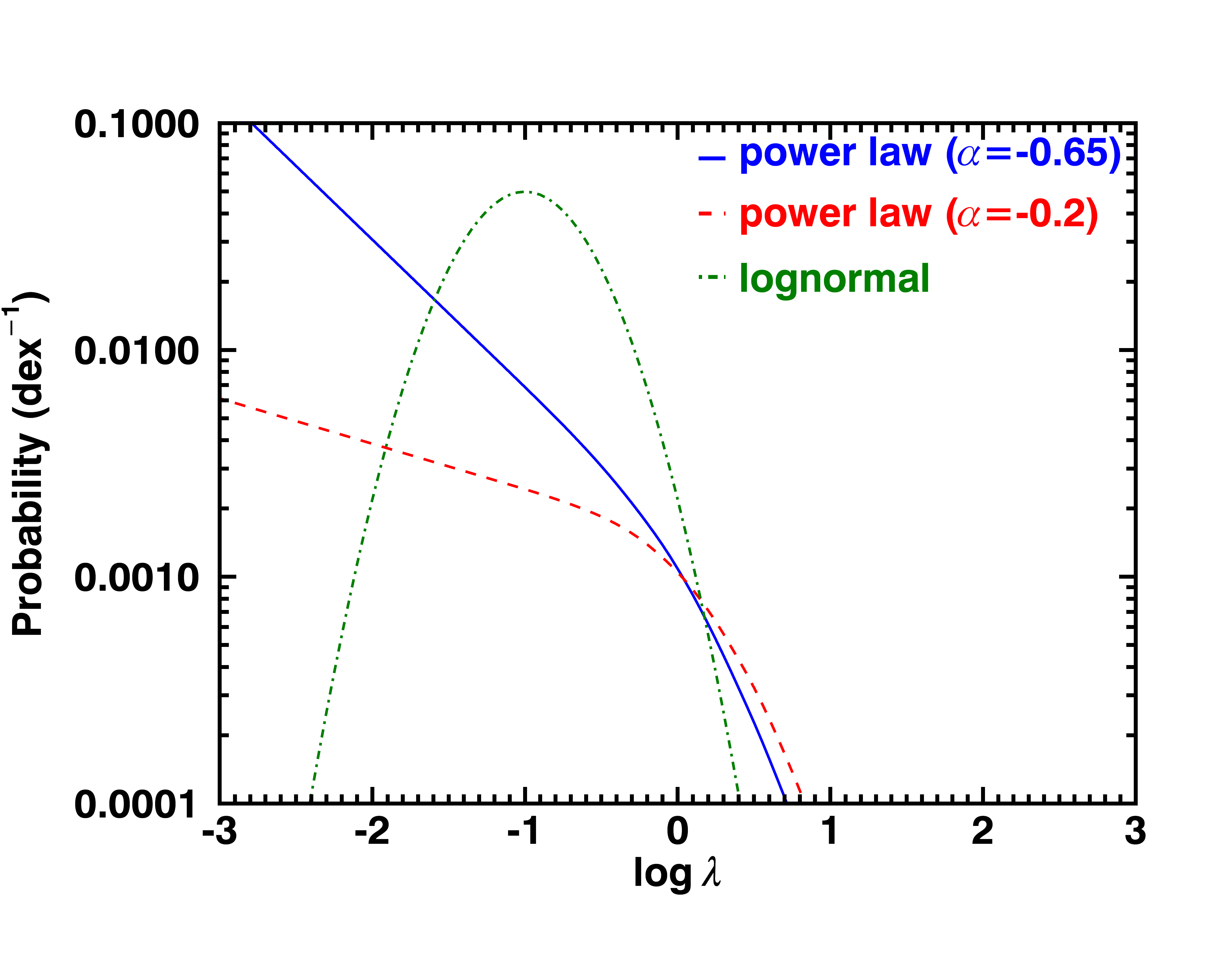}}
\caption{(left) Mean infrared luminosity from SF ($L_{\rm IR,SF}$)
  versus X-ray luminosity for X-ray selected AGNs over four redshift
  ranges. $L_{\rm IR,SF}$ is calculated over 8--1000~$\mu$m and has
  been corrected for the infrared luminosity from the AGN by
  decomposing the infrared SED into AGN and SF components. The
  infrared luminosity to SF rate conversion is from Kennicutt (1998)
  and the bolometric AGN luminosity ($L_{\rm AGN}$) is calculated from
  the X-ray luminosity using the luminosity dependent relation in
  Stern (2015). Predicted $L_{\rm IR,SF}$--$L_{\rm AGN}$ tracks for
  different model prescriptions of the Eddington-ratio distribution
  (as shown right) are plotted. Taken from Stanley et~al. (2015).}
\label{label1}
\end{figure*}

The first and third results imply a clear connection between AGN
activity and SF, in agreement with the similar shapes of the accretion
and SF rate densities; see Fig.~3. However, the second result appears
to imply a lack of a connection. How can the observed relationship
between AGN luminosity and SF luminosity depend on whether you
calculate the mean SF luminosity from the AGN population or the mean
AGN luminosity from the SF population? The likely answer lies in the
timescales of stability between AGN activity and SF (e.g.,\ Mullaney
et~al. 2012b; Gabor \& Bournaud 2013; Hickox et~al. 2014). AGNs are
known to vary by factors of several on short timescales (measured in
hours to years) and are likely to vary by orders of magnitude on
longer timescales (measured in thousands to millions of years) due to
changes in the Eddington ratio (i.e.,\ the mass accretion rate
relative to the black-hole mass). By comparison, SF is more stable on
short timescales and probably only varies significantly on timescales
of $>10^7$~years. Therefore, over the duration of a SF episode, the
observed AGN luminosity at a given time could differ from the mean AGN
luminosity by several orders of magnitude while the SF luminosity
would be comparatively stable. A consequence of this would be a
correlation between AGN and SF luminosity, if the AGN luminosity is
averaged as a function of SF luminosity (the more stable quantity),
and the absence of a correlation, if the SF luminosity is averaged as
a function of AGN luminosity (the more variable quantity). These basic
results are illustrated in the example study shown in Fig.~4. This
figure shows the mean SF luminosity as a function of AGN luminosity
and redshift for X-ray AGNs and also plots on the prediction for a
simple model whereby the AGN and SF luminosities are correlated on
long timescales but where the AGN luminosity can vary on short
timescales. The AGN variability prescription is taken from the
Eddington-ratio distribution of Aird et~al. (2012), also plotted on
Fig.~4. To first order, the good agreement between the simple model
and the data suggests that the basic hypothesis is correct. However,
we must be cautious about deriving more detailed conclusions from this
comparison since (1) we have only assumed a simple model and other
factors may play a role and (2) we have only been dealing with average
quantities.

\begin{figure}
\centerline{
\includegraphics[width=75mm]{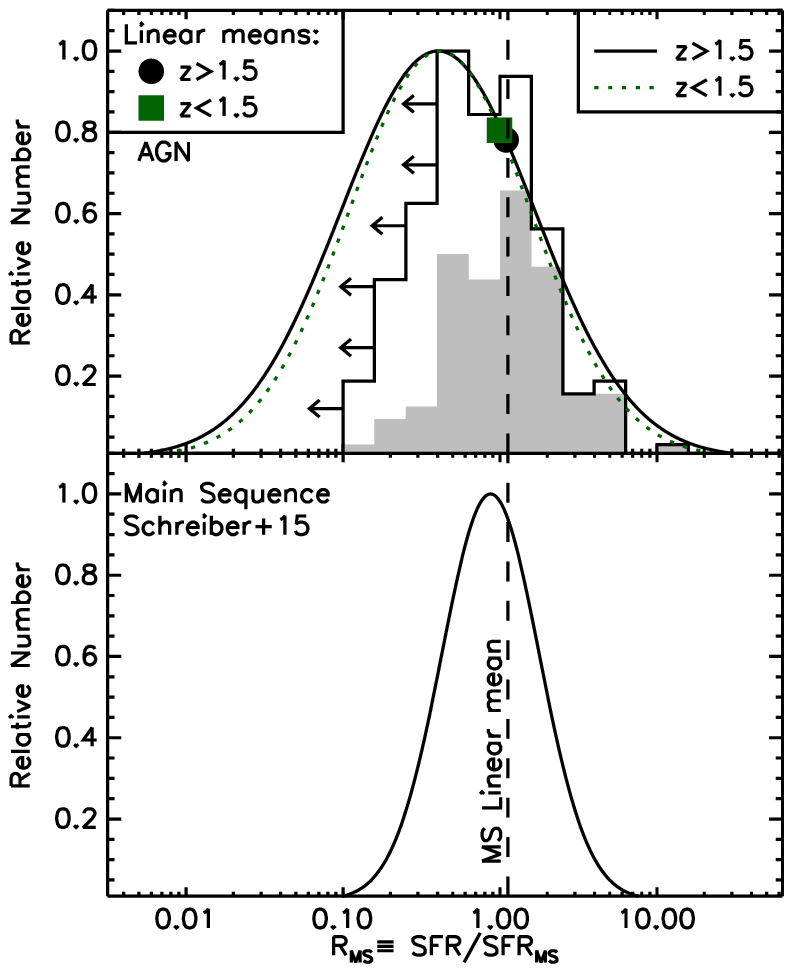}}
\caption{SF rate distributions for distant X-ray AGNs with ALMA
  constraints (top) and the SF galaxy population (bottom; from
  Schreiber et~al. 2015). The plotted distributions are the offset
  from the average of the SF galaxy population (i.e.,\ the average of
  the main sequence: SFR$_{\rm MS}$; e.g.,\ Schreiber et~al. 2015) and
  take account of the host-galaxy mass. The solid grey histogram shows
  the SF properties of the ALMA-detected AGNs, the open histogram
  shows the SF properties of the AGNs with ALMA upper limits, while
  the solid and dotted curves show the best-fitting log-normal
  distributions for X-ray AGNs at $z>1.5$ and $z<1.5$, respectively,
  taking account of the detections and upper limits. The solid circle
  and solid square show the mean values, which are in agreement with
  the mean of the SF galaxy population; however, the peaks of the
  distributions are shifted to lower SF luminosities, indicating that
  not all X-ray AGNs reside in SF galaxies. Taken from Mullaney
  et~al. (2015).}
\label{label1}
\end{figure}

The results presented so far have been based on mean quantities
because {\it Herschel} is not sensitive enough to individually detect
a large fraction of the distant X-ray AGN population. For a typical
log-normal distribution, the mean quantity will be biased towards the
brightest sources and will not be representative of a typical
source. Thankfully, improved SF luminosity constraints can be obtained
for individual X-ray AGNs using ALMA, which has significantly improved
sensitivity over {\it Herschel} at rest-frame far-infrared wavelengths
for AGNs and galaxies at $z>1$. The first results for X-ray AGNs from
ALMA are shown in Fig.~5. Despite the up-to an order of magnitude
greater sensitivity of ALMA over {\it Herschel}, the majority of the
X-ray AGNs are still undetected by ALMA (Mullaney
et~al. 2015). However, the ALMA upper limits are sensitive enough to
demonstrate that the distribution of specific SF rates for the X-ray
AGNs are inconsistent with that found for SF galaxies. The conclusion
from this work is therefore that not all distant X-ray AGNs reside in
SF galaxies, and that a substantial fraction (up-to $\approx$~50\%;
Mullaney et~al. 2015) are hosted in relatively quiescent galaxies with
low levels of SF; qualitatively similar results are found for X-ray
AGNs in the local Universe, using {\it Herschel} and the high-energy
(15--194~keV) {\it Swift}-BAT survey (Shimizu et~al. 2015). One
exciting explanation for the significant fraction of X-ray AGNs in
quiescent galaxies is that the SF has been shut off or suppressed by
AGN-driven outflows. However, it would be premature to reach that
conclusion from these data alone and other scenarios (e.g.,\ due to
the gas-inflow timescale, the accretion may proceed a long time after
the start and end of the SF episode) could also explain the data.

\section{Future prospects over the next decade}

The most significant new observatory on the immediate horizon for
X-ray--infrared synergy studies of AGNs is the {\it James Webb Space
  Telescope (JWST)} (Gardner et~al. 2006), with an expected launch in
2018. With high sensitivity, {\it Hubble Space Telescope}-like spatial
imaging quality, and spectroscopy (integrated and spatially resolved)
at infrared wavelengths of $\approx$~0.6--28~$\mu$m, {\it JWST}
promises to break new ground in both of the science themes explored
here. From the combination of {\it JWST} and {\it XMM-Newton}, {\it
  Chandra}, and {\it NuSTAR}, we should anticipate great advances in
measuring the combined infrared and X-ray properties of AGNs and in
completing the overall census of AGN activity. While the mid-infrared
waveband does not probe the peak of the dust-obscured SF, {\it JWST}
will also improve our constraints on the SF properties of X-ray AGNs
by better identifying and constraining the AGN component either
through high-spatial resolution imaging or spectroscopy. At the end of
the next decade, {\it Athena} (Nandra et~al. 2013) will launch
(expected launch in 2028) and usher in a new era of X-ray
astronomy. With unsurpassed X-ray spectral resolution and high
sensitivity over a large field of view, {\it Athena} will dramatically
push forward our census of the AGN population and our understanding of
the connection of AGN activity with SF. However, {\it Athena} is
unlikely to overlap with {\it JWST} (which has an expected lifetime of
$\approx$~5--10~yrs) and, therefore, to optimise the impact and
discovery potential, the current suite of X-ray observatories ({\it
  XMM-Newton}; {\it Chandra}; {\it NuSTAR}) will need to be
operational over the lifetime of {\it JWST}.


\acknowledgements I thank the substantial number of long-suffering
collaborators who have worked with me over the last decade in Durham
and beyond for their inspiring scientific discussions and research. I
also thank the referee for a concise and constructive report. I
acknowledge the Science and Technology Facilities Council (STFC) for
support through grant ST/L00075X/1.


\end{document}